\documentclass[10pt]{article}
\usepackage{times}
\usepackage{epsfig}
\title{{\bf Ward-Takahashi Identity on the Light Front$^*$
\footnote{Submitted in Few Body Systems}}}
\author{ \\
\\
\\
\\  H. W. L. Naus$^a$, J. P. B. C. de Melo$^b$ and T. Frederico$^c$
\\
\\
 $^a$ {\em Institute for Theoretical Physics, University of Hannover}\\
{\em Appelstr. 2, 30167 Hannover, Germany}\\
$^b$ {\em Instituto de F\'\i sica, Universidade de S\~ao Paulo}\\
{\em 01498-970 S\~ao Paulo, S\~ao Paulo, Brazil}\\
$^c$ {\em Dep. de F\'\i sica, Instituto Tecnol\'ogico da Aeron\'autica,}\\
{\em Centro T\'ecnico Aeroespacial,  12.228-900 S\~ao Jos\'e dos Campos,
S\~ao Paulo, Brazil}\\
 \\
\\
\\
\\}
\date{\today \
\\
 \begin{abstract}
The Ward-Takahashi identity, reflecting local gauge invariance,
is perturbatively verified for a boson model
in light front field theory. 
A careful integration over the light front energy, corresponding
to exactly taking into account pair terms,
which are the contributions of the zero longitudinal
momentum mode, is crucial to
obtain this result. 
Furthermore, the one-loop boson form factors are calculated
for arbitrary off-shell momenta.\\
\bigskip
\bigskip
*{\it Dedicated to Prof. John A. Tjon on the occasion of his 60th birthday.}
\end{abstract}}
\begin{document}
\maketitle
\setcounter{equation}{0}
\pagebreak
\baselineskip=24.0pt
\section{Introduction}
Ward-Takahashi (WT) identities \cite{WA, TA} are among the most interesting
consequences of local $U(1)$ gauge invariance in quantum field theory.
It appears therefore surprising that they have not
been addressed in light front field theory.  One reason may be that
a formal proof \cite{WA, TA, IZ}, exploiting equal time canonical
commutation relations, cannot be taken over to the light front. Front
form dynamics can only be described as a constrained system and not
be quantized canonically. In fermionic theories, for example, some
components of the spinor fields are not independent but
necessarily obey constraint equations. These ``bad" components
also enter in the electromagnetic operators, which, in turn,
lead to bad currents. Recall the simplest WT identity: the relation
between electromagnetic vertex  and propagator. Because current and
electromagnetic vertex are evidently related, the question arises
whether the WT identity actually holds on the light front.

Instead of a formal and more general approach to this problem, we
first study a simple model using perturbation theory. We explicitly
calculate vertex function and self-energy of a charged
boson; its structure arises from its interaction with
a neutral and a charged boson. Our approach is based
on a series of papers by Chang, Root and Yan \cite{CR, CH, YA1, YA2},
where it was demonstrated
that light front perturbative field theory amounts to using
the Feynman rules and then {\it first}
integrating over the light front energy, $p^-$, in momentum integrals.
It can be compared to (re-)deriving time ordered perturbation
theory by integration over the energy variable $p^0 = E$.
At this point several remarks are in order. First, special care
is needed in the actual $p^-$ integration since a too naive
approach yields wrong answers \cite{YA2}. We will come back to
this point below. Second, this simple recipe cannot readily
be applied to theories with massless particles and/or gauge theories.
Since we consider massive bosons and we do not include
radiative corrections this poses no restriction for our work.
Finally, in recent years the modes of vanishing light front
momentum $p^+$ have attracted
much attention (see, e.g. \cite{PNK} and references therein).
It is a priori not clear whether the recipe also works in this
zero mode sector.
We speculate, however, that the
careful integration mentioned above, indeed renders this problem for
massive virtual particles.

Recently, also the role of the pair (creating) terms has been
discussed for the same boson model \cite{Melo}. It was shown that
the pair terms survive even in the limit of zero light front
momentum $p^+$ and that they are necessary to obtain a covariant
current. Here we extend this work to general off-shell momenta,
consider WT identities and confirm the relevance
of the pair terms. Furthermore, all components of the
electromagnetic vertex operator in an arbitrary Lorentz-frame are
constructed. Obviously, our calculation produces the on-shell
form factor, its off-shell extension, as well as the
additional off-shell form factor. In refs. \cite{Naus, Tjon} similar
issues for the nucleon, however using a meson field theory in the
instant form, have been investigated.

In this paper, we again employ the 
``dislocation of pole integration'' developed in \cite{Melo}. On the
other hand, in contrast to the latter work we start from
one basic integral which, after regularization, can be completely performed.
Thus after the  $p^-$ integration, the
integrations over the other components of $p$ are carried out. 
In this way we can immediately calculate all components of
the boson current, demonstrating that for this simple model
the term ``bad component'' is somewhat artificial.
Since we are interested in WT identities gauge invariant regularization
and renormalization procedures are preferred, ruling out
excessively naive momentum cut-offs. 

\section{Basic Techniques}
We first develop the basic calculational techniques.
Consider the following integral in light front variables
\begin{equation}
I_1(M^2)= \int d^4 p \, \frac{1}{p^2-M^2+i\epsilon} = \frac{1}{2}
\int d^2 p_{\perp} \, \int d p^+ \, \int d p^- \,
\frac{1}{p^+ p^- - p_{\perp}^2 -M^2 +i \epsilon} \, ,
\end{equation}
where one needs to integrate first over the light front energy $p^-$. 
Recall that a naive light front integration immediately would
yield zero \cite{YA2}, obviously the wrong answer. Furthermore,
note that this integral is divergent. In order to regularize
the integral in the ultraviolet domain we introduce
a regulator $M_R$ \cite{Pau}: 
\begin{equation}
I_1^{reg}(M^2, M_R^2) =  \frac{1}{2}
\int d^2 p_{\perp} \, \int d p^+ \, \int d p^- \,
\frac{1}{p^+ p^- - p_{\perp}^2 -M^2 +i \epsilon} 
\left(\frac{-M_R^2}{p^+ p^- - p_{\perp}^2 -M_R^2 +i \epsilon}\right)^2 \, .
\end{equation}
At the end of the day, the regulator is supposed to approach infinity.
Let us introduce the notation $\tilde{M}^2 = M^2 + p_{\perp}^2$
and $\tilde{M}_R^2 = M_R^2 + p_{\perp}^2$. To prepare the $p^-$
integration one separates a factor $1/p^+$, e.g.
\begin{equation}
\frac{1}{p^+ p^- - p_{\perp}^2 -M^2 +i \epsilon} =
\frac{1}{p^+} \, \frac{1}{ p^- - \frac{\tilde{M}^2 -i \epsilon}{p^+}} \, .
\end{equation}
Clearly, this procedure may cause problems for $p^+=0$, the
zero momentum mode. In order to regulate this infrared singularity
we dislocate the other poles \cite{Melo}:
\begin{equation}
\frac{1}{p^+ p^- - p_{\perp}^2 -M_R^2 +i \epsilon} \rightarrow
\frac{1}{p^+ \pm \delta} \,
\frac{1}{ p^- - \frac{\tilde{M}_R^2 -i \epsilon}{p^+ \, \pm \, \delta}} \, ,
\end{equation}
with a small parameter $\delta$. Then the regularized
integral reads
\begin{equation}
I_1^{reg}(M^2, M_R^2) =  \frac{1}{2}
\int d^2 p_{\perp} \, \int d p^+ \,
\frac{1}{p^+(p^++\delta)(p^+-\delta)}
\int d p^- \,
\frac{1}{p^- - \frac{\tilde{M}^2 -i \epsilon}{p^+}} \,
\frac{M_R^2}{p^- - \frac{\tilde{M}_R^2 -i \epsilon}{p^++\delta}} \,
\frac{M_R^2}{p^- - \frac{\tilde{M}_R^2 -i \epsilon}{p^+-\delta}}  \, .
\end{equation}
With a modest amount of foresight we do not include $\delta$ as an
additional argument of $I_1^{reg}$.
At this point we integrate over $p^-$; the result can be written as
\begin{equation}
I_1^{reg}(M^2, M_R^2) = i \pi M_R^4
\int d^2 p_{\perp} \, \frac{1}{\tilde{M}_R^2 \, \delta} \,
\int_0^{\delta} d p^+ \,
\frac{p^+-\delta}{p^+(M_R^2-M^2) + \tilde{M}^2 \delta} \,.
\end{equation}
Doing the $p^+$ integration yields 
\begin{equation}
I_1^{reg}(M^2, M_R^2) = \frac{i \pi M_R^4}{M_R^2-M^2}
\int d^2 p_{\perp} \, \frac{1}{\tilde{M}_R^2} \left(
1 - \frac{\tilde{M}_R^2}{\tilde{M}_R^2-M^2}
\ln \frac{\tilde{M}_R^2}{\tilde{M}^2} \right) \, .
\end{equation}
Note that this expression does not depend anymore on the infrared regulator
$\delta$, a forteriori justifying the procedure. 
The ``dislocation of the pole'' is  vanishing small, thus in fact
the contribution to the integration arises from the momentum $p^+$ near zero.
This zero momentum mode can be
interpreted as a pair term for finite $\delta$. 
Finally, we
integrate over the transverse momenta and we obtain
\begin{equation}
I_1^{reg}(M^2, M_R^2) = i \pi^2 M_R^4 \left(
\frac{M^2}{(M^2 - M_R^2)^2} \ln \frac{M_R^2}{M^2} 
+ \frac{1}{M^2-M_R^2}\right) \, .
\end{equation}
Here the quadratic divergence explicitly shows up.

This result can indeed be considered as basic because other
integrals, for example,
\begin{equation}
I_2^{reg}(M^2, M_R^2) =  \frac{1}{2}
\int d^2 p_{\perp} \, \int d p^+ \,  \int d p^- \,
\left(\frac{1}{p^+ p^- - p_{\perp}^2 -M^2 +i \epsilon}\right)^2 
\frac{-M_R^2}{p^+ p^- - p_{\perp}^2 -M_R^2 +i \epsilon} \, ,
\end{equation}
can readily be obtained from it.
Note that we have already regularized $I_2$; 
now we immediately get
\begin{equation}
I_2^{reg}(M^2, M_R^2) = - \frac{M_R^2}{M^4} \,
I_1^{reg}(M_R^2, M^2)  = i \pi^2 M_R^2 \left(
\frac{M_R^2}{(M^2 - M_R^2)^2} \ln \frac{M_R^2}{M^2} 
+ \frac{1}{M^2-M_R^2}\right) \, .
\end{equation}
Indeed $I_2$ has a logarithmic divergence.
Herewith, we also find the convergent integral
\begin{eqnarray}
I_3(M^2) &=&  \frac{1}{2}
\int d^2 p_{\perp} \, \int d p^+ \, \int d p^- \,
\left(\frac{1}{p^+ p^- - p_{\perp}^2 -M^2 +i \epsilon}\right)^3 \nonumber \\
&=& \lim_{M_R^2 \rightarrow \infty} \, \frac{1}{2} \,
\frac{\partial}{\partial M^2} \, I_2^{reg}(M^2, M_R^2) = 
\frac{-i \pi ^2}{2 M^2} \, ,
\end{eqnarray}
which is in agreement with the result of ref. \cite{YA2}.

Finally, we consider the regularized integrals
\begin{equation}
I_k^{reg}(q, M^2, M_R^2) =  \frac{1}{2}
\int d^2 p_{\perp} \, \int d p^+ \,  \int d p^- \,
\left(\frac{1}{p^2 +2 p q -M^2 +i \epsilon}\right)^k 
\left(\frac{-M_R^2}{p^2 + 2 p q -M_R^2 +i \epsilon}\right)^{3-k} \, ,
\end{equation}
with $k=1, 2 \, .$ Since shifts in the momenta are allowed 
in these convergent expressions, we easily verify
\begin{equation}
I_k^{reg}(q, M^2, M_R^2) = \left(\frac{M_R^2}{M_R^2+q^2}\right)^{3-k}
I_k^{reg} (M^2 + q^2, M_R^2 + q^2)  \,.
\end{equation}
Similarly, we obtain the analogous expression for the convergent integral
\begin{equation}
I_3(q, M^2) =  \frac{1}{2}
\int d^2 p_{\perp} \, \int d p^+ \, \int d p^- \,
\left(\frac{1}{p^2 +2 p q -M^2 +i \epsilon}\right)^3 =
\frac{-i \pi ^2}{2 (M^2+q^2)} \, .
\end{equation}

We conclude that despite the first integration over $p^-$ all
the results are manifestly covariant. Again we confirm the
connection between a careful $ p^-$ integration, in this case
via dislocating coinciding poles, and covariance \cite{Melo}.
Of course, the use of a covariant regularization
procedure is also crucial in this respect.

\section{Vertex Function, Self-Energy and Ward-Takahashi Identity}
With the results obtained so far, the calculation of the vertex
function and self-energy of the composite boson is
standard. At this point, light front aspects do not explicitly
show up anymore since they have already be taken care of in the
evaluation of the relevant integrals. Thus the whole calculation
now appears covariant.

The one-loop irreducible vertex function for the boson model \cite{Melo} reads 
\begin{equation}
\Gamma_\mu = e \left[ (p+p')_\mu +  \Lambda_\mu \right] \, ,
\end{equation}
with the one-loop ($O(g^2)$) vertex correction
\begin{equation}
\Lambda_\mu =  - g^2 \int \frac{d^4k}{(2\pi)^4}\,
\frac{i^3 (2k - p -p')_\mu}{((k-p)^2-m^2+i\epsilon)
((k-p')^2-m^2+i\epsilon) (k^2-m^2+i\epsilon)} \, .
\end{equation}
The masses of charged and neutral constituents are taken to be
equal and are denoted by $m$. We do not consider the external boson
to be on its mass shell yet, thus in general $p^2 \ne M^2, p'^2 \ne M^2$.
Combining the denominators using the Feynman trick, we get
\begin{equation}
\Lambda_\mu = 2i g^2 \int_0^1 dx \, \int_0^x dy \,  \int \frac{d^4k}{(2\pi)^4}\,
\frac{(2k - p -p')_\mu}{(k^2+2 k \xi -m_0^2+i\epsilon)^3}\, ,
\end{equation}
where 
$ \xi = y(p'-p) - xp'\,$  and
$m_0^2 = m^2 + (p'^2-p^2)y - p'^2x \,$. Below we will also use
$q=p'-p\,$.
The momentum integral is convergent and using the results
derived above we obtain
\begin{equation}
\Lambda_\mu = \frac{i^2g ^2 \pi^2}{(2\pi)^4} \int_0^1 dx \, \int_0^x dy \,  
\frac{( p +p')_\mu +2 \xi_\mu}{\xi^2 +m_0^2-i\epsilon}\, .
\end{equation}
After rearranging we identify the form factors $F_1$ and $F_2$
\begin{equation}
\Gamma_\mu = e\left[F_1(q^2, p^2, p'^2) 
( p'+p)_\mu + F_2(q^2, p^2, p'^2) (p'-p)_\mu\right] \, ,
\end{equation}
where $F_1(q^2, p^2, p'^2) =1 +
f_1(q^2, p^2, p'^2), \,
F_2(q^2, p^2, p'^2) = f_2(q^2, p^2, p'^2)$ and
\begin{eqnarray}
f_1 &=& \frac{g^2\pi^2}{(2\pi)^4}\int_0^1 dx \, \int_0^x dy \,
\frac{x-1}{(p'^2-p^2)y-p'^2x+y^2q^2+(p^2-q^2-p'^2)xy+x^2p'^2+m^2-i\epsilon}
\, ,\nonumber \\
f_2 &=& \frac{g^2\pi^2}{(2\pi)^4} \int_0^1 dx \, \int_0^x dy \,
\frac{x-2y}{(p'^2-p^2)y-p'^2x+y^2q^2+(p^2-q^2-p'^2)xy+x^2p'^2+m^2-i\epsilon}\, .
\end{eqnarray}

The one-loop ($O(g^2)$) self-energy is given by
\begin{eqnarray}
-i \Sigma (p^2) &=&  g^2 \int \frac{d^4 k}{(2 \pi)^4} 
\frac{i}{(k+\frac{p}{2})^2-m^2+i\epsilon} \,
\frac{i}{(k-\frac{p}{2})^2-m^2+i\epsilon}  \nonumber \\
&=&   - g^2\int_0^1 dx \,  \int \frac{d^4k}{(2 \pi)^4}
\frac{1}{(k^2 +2 k \eta -\mu^2+i\epsilon)^2} \, ,
\end{eqnarray}
where
$\eta =   (x-\frac{1}{2})p$ and $ \mu^2 = m^2 -\frac{1}{4}p^2 \,$.
This momentum integral is divergent and we regularize
as described above and get
\begin{equation}
-i \Sigma (p^2) = \frac{-g^2}{(2\pi)^4} \int_0^1 dx\,
I_2^{reg}(\eta, \mu^2, M_R^2)  =
\frac{-g^2 }{(2\pi)^4}\int_0^1 dx\, \frac{M_R^2}{M_R^2+\eta^2}
I_2^{reg}(\mu^2+\eta^2, M_R^2 +\eta^2) \, .
\end{equation}
For large regulator $M_R^2$ we obtain
\begin{equation}
\Sigma (p^2)= \frac{\pi^2 g^2}{(2\pi)^4} \int_0^1 dx  \left[
\ln\left(\frac{M_R^2}{\mu^2 + \eta^2}\right) - 1 \right] + \chi \, ,
\end{equation}
where $\chi$ denote terms vanishing in the limit $M_R \rightarrow \infty\, .$

At this point, {\it i.e.}, before renormalization, we may 
already verify the WT identity 
\begin{equation}
(p'-p)^\mu \Gamma_\mu = e \left[ \Delta^{-1}(p'^2) - \Delta^{-1}(p^2) \right] \, ,
\end{equation}
with the full propagator $\Delta$. In terms of self-energy
and vertex correction it reads
\begin{equation}
(p'-p)^\mu \Lambda_\mu = \Sigma (p^2) - \Sigma (p'^2) \, .
\end{equation}
For the lhs we find
\begin{eqnarray}
(p'-p)^\mu \Lambda_\mu &=& 
-\frac{g ^2 \pi^2}{(2\pi)^4} \int_0^1 dx \, \int_0^x dy \,  
\frac{(1-x)(p'^2-p^2)+(2y-x)q^2}{\xi^2 +m_0^2-i\epsilon} \nonumber \\
&=& \frac{g ^2 \pi^2}{(2\pi)^4} \int_0^1 dx \,   
\ln \left(\frac{m^2-p'^2x+x^2p'^2-i\epsilon}
{m^2-p^2x+x^2p^2-i\epsilon}\right) \, .
\end{eqnarray}
In the rhs the infinities indeed cancel, which allows us to
take the limit $\displaystyle M_R \rightarrow \infty$. Then we obtain
\begin{equation}
\Sigma(p^2) -\Sigma(p'^2) = 
\frac{g ^2 \pi^2}{(2\pi)^4} \int_0^1 dx \,   
\ln \left(\frac{m^2-p'^2x+x^2p'^2-i\epsilon}
{m^2-p^2x+x^2p^2-i\epsilon}\right)\, .
\end{equation}
Thus the WT identity holds.

\section{Renormalization and Form Factors}
In order to extract observables, e.g. on-shell form factors,
perturbative renormalization is nevertheless necessary. 
Choosing the on-shell subtraction scheme \cite{IZ, Naus, Tjon} amounts to
replacing the divergent self-energy:
\begin{equation}
\Sigma(p^2) \rightarrow  \Sigma_R(p^2) =
\lim_{M_R \rightarrow \infty}
\left[\Sigma(p^2) -
\Sigma(M^2) - (p^2 - M^2) \Sigma'(M^2)\right] \, .
\end{equation}
Since $\Sigma_R(M^2) = 0$ and $\Sigma'_R(M^2) = 0$, the
renormalized propagator has a pole at the physical mass,
$p^2 = M^2$, with residue 1. These subtractions correspond to an
infinite mass renormalization and a finite wave function
renormalization. The latter also induces a finite renormalization of 
the vertex function
\begin{equation}
\Gamma_\mu \rightarrow  \Gamma_\mu^R =e\left[F_1^R(q^2, p^2, p'^2)(p+p')_\mu +
F_2^R(q^2, p^2, p'^2)(p-p')_\mu\right] =
\Gamma_\mu +(p+p')_\mu \ \Sigma'(M^2) \, ,
\end{equation}
which implies
\begin{equation}
F_1^R  = F_1 -
\frac{\pi^2 g^2}{(2\pi)^4}\int_0^1 dx \, \frac{x(x-1)}{M^2x(x-1)+m^2} \,.
\end{equation}
The second form factor does not change, $F_2^R=F_2$. One easily
checks that the WT identity is also valid for the renormalized
quantities.
The reducible vertex function is related to
the irreducible one via 
\begin{equation}
\Delta_0(p'^2) \Gamma_\mu^{red} \Delta_0(p^2) =
\Delta(p'^2) \Gamma_\mu^R \Delta_(p^2) \, ,
\end{equation}
with the free propagator $\Delta_0(p^2) = (p^2 - M^2)^{-1}.$ It also can
be written in terms of two form factors 
\begin{equation}
\Gamma_\mu^{red} = e\left[g_1(q^2, p^2, p'^2)(p+p')_\mu +
g_2(q^2, p^2, p'^2)(p-p')_\mu\right] \, .
\end{equation}
Perturbatively it follows that
\begin{equation}
g_1(q^2, p^2, p'^2) = F_1^R(q^2, p^2, p'^2) + \Sigma_R(p'^2) \Delta_0(p'^2)
+ \Sigma_R(p^2) \Delta_0(p^2) \, ,
\end{equation}
and $g_2 = F_2^R$ . In the on-shell limit the additional terms vanish:
$\displaystyle \lim_{p^2 \rightarrow M^2} \Sigma_R(p^2) \Delta_0(p^2) =0$.
Particularly interesting is the half off-shell case at the photon
point. Then the WT identity implies  
$g_1(0, p^2, M^2) = 1$, which means that even in the half off-shell case
the interpretation of $g_1$ as charge form factor makes sense.
Using the expressions given above, it is straightforward to analytically
check this relation for our model.

Let us now consider the
form factors of the reducible renormalized vertex function.
We restrict ourselves to the half off-shell case,  $p'^2=M^2$,
and take $q^2 < 0$ and $p^2 < 4m^2$.
In this way, no poles in the integrand are present and,
consequently, no imaginary parts appear due to real two-boson
production. The integrals over the Feynman parameters
are evaluated numerically.
We define our units via $M^2=1$; then we take $m^2=1.1,\, g^2=0.1\,$.
In Figs. (1) and (2) we respectively present $g_1$ and $g_2$
as a function of $-q^2$ for
different off-shell values of $p^2.$  It is established that
$g_1(0, p^2, M^2) = 1$ and $g_2(q^2, M^2, M^2) =0$ -only one
form factor is present in the on-shell limit. Both form factors
reflect appreciable off-shell variations. Thus also perturbative
light front field theory generates these effects.

\section{Conclusion and Outlook}
We have studied a composite charged boson in the context of a simple
light front field theory. The electromagnetic vertex and self-energy
of the boson have been perturbatively calculated for arbitrary off-shell
momenta. Up to second order in the coupling, corresponding to the one-loop
approximation, 
we have explicitly verified the WT identity on the light front.
The  accurate treatment of the zero momentum mode is crucial for
this result.

In the nearby future we plan to address similar light front problems
for the pion, with fermionic constituents, and   
for a spin 1/2 particle like the nucleon. In case these studies
also indicate the validity of the WT identity, it would
be interesting trying to establish general, nonperturbative
WT identities in light front field theory.

\section*{Acknowledgements}
This work was supported in part by
Funda\c c\~ao Coordena\c c\~ao 
de Aperfei\c coamento de Pessoal de N\'\i vel Superior and 
the Deutscher Akademischer Austauschdienst
(Probral/CAPES/DAAD project 015/95).
It was also supported by 
the Brazilian agencies CNPq and FAPESP. 
H. W. L. N. acknowledges the hospitality of the
Instituto Tecnol\'ogico da Aeron\'autica, 
Centro T\'ecnico Aeroespacial in S\~ao Jos\'e dos Campos.
J. P. B. C. M. acknowledges the hospitality of the
Institute for Theoretical Physics, University of Hannover.
The  authors thank A. C. Kalloniatis and
P. U. Sauer for useful discussions and a critical
reading of the manuscript.

\newpage

\begin{figure}[h]
\vspace{15.0cm}
\includegraphics{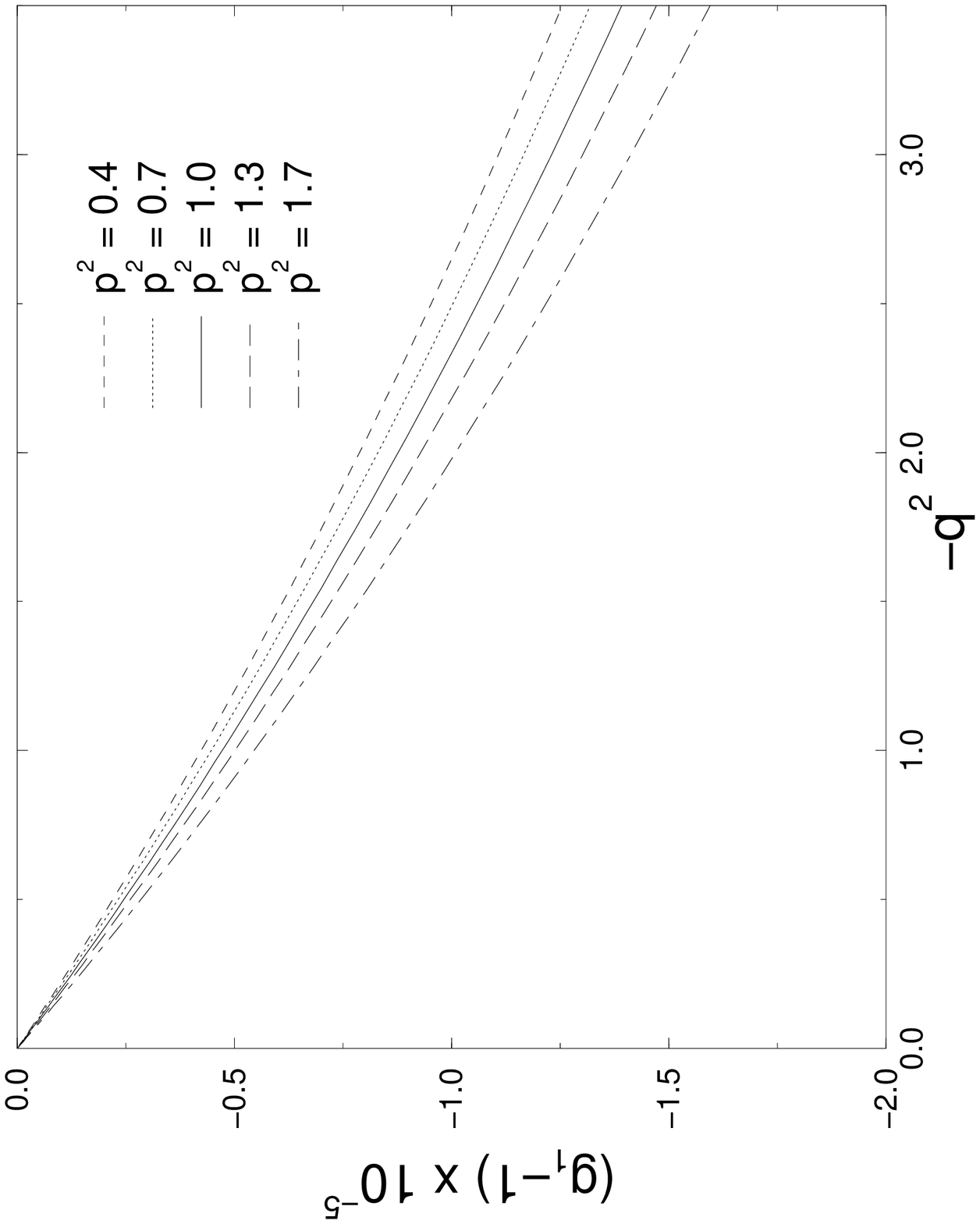}
\caption{
the form factor $g_1\,$. Note that we have plotted ($g_1$ - 1).}
\end{figure}

\newpage

\begin{figure}[h]
\vspace{15.0cm}
\includegraphics{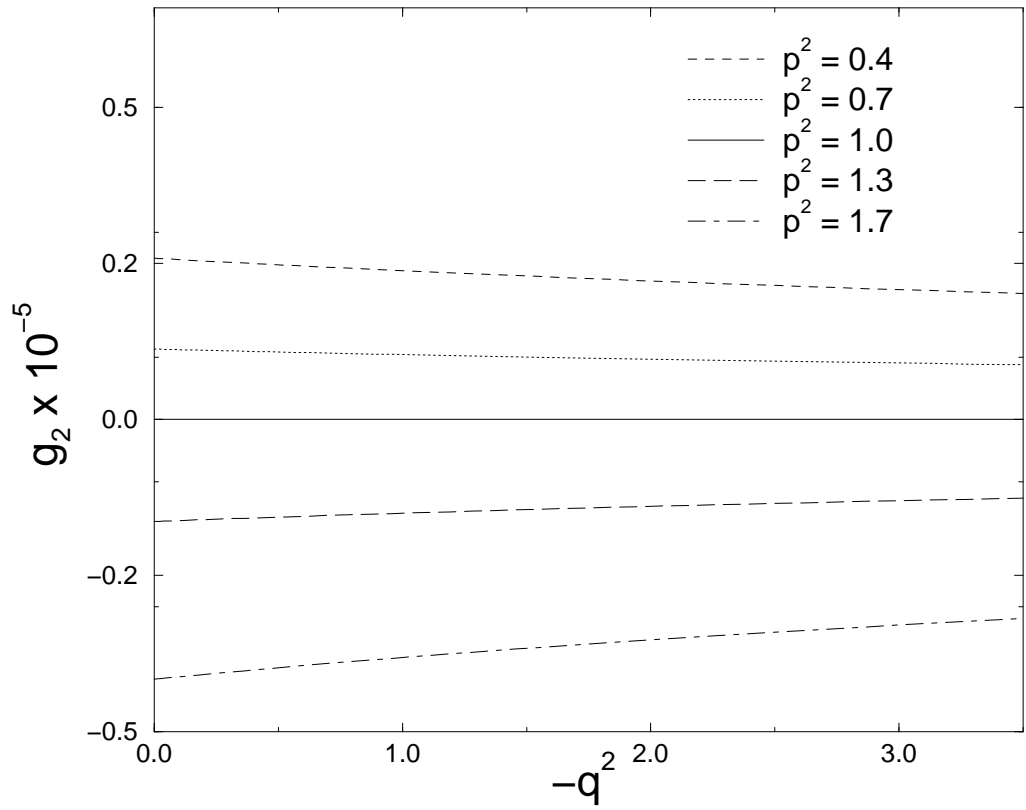}
\caption{
the form factor $g_2$.}
\end{figure}

\end{document}